# Mapping computational thinking mindsets between educational levels with cognitive network science

M. Stella[1], A. Kapuza[2], C. Cramer[3] and S. Uzzo[4]

1 Complex Science Consulting, Via Amilcare Foscarini 2, 73100 Lecce, Italy
2 National Research University Higher School of Economics, Myasnitskaya 20, 101000 Moscow, Russian Federation
3 Woods Hole Institute, Woods Hole, MA, USA
4 New York Hall of Science, Corona, NY, USA

*Computational thinking is a way of reasoning about the world in terms of data. This mindset channels number crunching toward an ambition to discover knowledge through logic, models and simulations. Here we show how computational cognitive science can be used to reconstruct and analyse the structure of computational thinking mindsets (forma mentis in Latin) through complex networks. As a case study, we investigate cognitive networks tied to key concepts of computational thinking provided by: (i) 159 high school students enrolled in a science curriculum and (ii) 59 researchers in complex systems and simulations. Researchers' reconstructed forma mentis highlighted a positive mindset about scientific modelling, semantically framing data and simulations as ways of discovering nature. Students correctly identified different aspects of logic reasoning but perceived "computation" as a distressing, anxiety-eliciting task, framed with math jargon and lacking links to real-world discovery. Students' mindsets around "data", "model" and "simulations" critically revealed no awareness of numerical modelling as a way for understanding the world. Our findings provide evidence of a crippled computational thinking mindset in students, who acquire mathematical skills that are not channelled toward real-world discovery through coding. This unlinked knowledge ends up being perceived as distressing number-crunching expertise with no relevant outcome. The virtuous mindset of researchers reported here indicates that computational thinking can be restored by training students specifically in coding, modelling and simulations in relation to discovering nature. Our approach opens innovative ways for quantifying computational thinking and enhancing its development through mindset reconstruction.*

# Introduction

Computational thinking is a way of thinking about the world in terms of data. This mindset recasts ideas or problems as quantification tasks, in which data, theory, models and simulations are combined creatively to achieve quantitative answers (Weintrop et al. 2016; Cramer et al. 2018). Computational thinking skills include: (i) defining a quantifiable problem (e.g. how quickly can a virus spread?), (ii) using logic reasoning for mapping the problem in a computable model of reality (e.g. using data and theory for simulating how individuals can get infected), (iii) computing the answer (e.g., each infected person can infect three more, on average), and (iv) interpreting the results (e.g., the virus spreads faster than the flu). Computational thinking is a challenging way of understanding our world, requiring creativity, expertise, data availability and knowledge. Computers are not essential for the easier tasks, but prove invaluable for analysing large volumes of data or testing theoretical predictions by coding simulations (e.g., NASA's picture of a black hole required processing 5 petabytes of data).

As a way of perceiving and linking different elements of reality, i.e. a mindset, a definition of computational thinking grounded in cognitive science remains an open question. Identifying the presence of this mindset is key in our society, which increasingly depends on data processing (e.g., for simulating virus spreading) and requires more and more computational skills that are in high demand in the job market (OECD, 2019).

The attitude toward the subject is an important part of mindsets. Studies show that while mathematical knowledge could be more highly related to short-term outcomes, such as achievement in a statistics course (Beurze et al., 2013; Chiesi & Primi, 2010), the attitudes toward statistics are related to long-term outcomes, such as dissertation grade (Vanhoof et al., 2006). However, other studies show that some components of these attitudes have a strong relationship even with short-term results (Dempster & McCorry, 2009; Emmioglu & Capa-Aydin, 2012). More complex models suggest that components of attitudes, such as interest and value, could affect others, such as effort and expectancy for success, that, in turn, are directly related to subject achievements (Ivaniushina et al., 2016; Ramirez et al., 2012).

In general, computational thinking is "the thought processes involved in formulating a problem and expressing its solution(s) in such a way that a computer—human or machine—can effectively carry out" (Wing, 2014). There is no consensus about an exact definition, but in all approaches it is broader than counting skills and also includes the ability to think logically and abstractly and use problem-solving skills (see Wing, 2006; Shute et al., 2017). Computational thinking plays an important role in the whole of STEM education and beyond (DePryck, 2016; Weese & Feldhausen, 2017; Weintrop et al., 2016)

This paper moves a step forward toward investigating the presence and structure of a computational thinking mindset by using the framework of forma mentis networks (FMNs), i.e. knowledge graphs that map conceptual knowledge and sentiment as perceived by individuals (Stella et al. 2019). In computational cognitive science, knowledge graphs map conceptual associations (Popping, 2003), including a wide variety of memory recall or semantic links between concepts and information (e.g. "virus" evoking the memory of "disease", see De Deyne et al. 2016). The structure of knowledge graphs has been shown to influence many cognitive processes (Siew et al. 2019), such as writing styles (Amancio, 2015), stance detection (Stella et al,. 2018) and creativity levels (Kenett et al., 2014). In education research, maps of conceptual associations revealed important insights in the way students acquire and structure their knowledge (Koponen and Kokkonen, 2014; Cramer et al. 2018; Siew 2019). In addition to knowledge structure, the positive/negative sentiment attributed to each concept also plays a fundamental role in thought processes (Kuperman et al. 2013). Forma mentis (Latin for "mindset") networks combine these elements in order to reconstruct the semantic frame (Fillmore, 2006), or meaning, and emotions attributed by individuals to concepts, thus reconstructing a networked model of their mindset (Stella et al. 2019).

In this study we examine the representation of key concepts of data science and computational thinking in experts' and students' mindsets. Based on the literature, several key concepts were chosen. Firstly, such general concepts as "computer science" and "computers" were used. Second, one of the important components of computational thinking – "logical reasoning" – was considered (Csizmadia et al., 2015; Shute et al., 2017; Vivian et al., 2014). Then, based on computational thinking in mathematics and science taxonomy (Weintrop et al., 2016), "model", "computation", "simulation", "data" and "code" were highlighted.

Herein we focus on detecting traces of computational thinking patterns in the mindsets of two populations: (i) 159 high school students enrolled in a STEM-focused curriculum, and (ii) 59 STEM researchers working on modelling complex systems.

STEM experts' mindset is seen to be fully aligned with key aspects of computational thinking as identified by previous studies, namely awareness about logic reasoning; a positive attitude toward data, simulations and models; and the ability to frame coding as a tool for achieving new knowledge. Students' mindset was quite different. Although they are aware of logic reasoning, students framed "model" as vaguely related to computing. They also completely missed any link between "simulation" and computation, which was perceived as an anxiety-eliciting construct. We interpret this lack of perception as a partially undeveloped computational mindset, possessing fundamental tools for logic reasoning but lacking the "sense of purpose" that psychologically drives computational thinking with a need to solve a challenge for understanding our world. Restoring this purpose in students, as it is in researchers, is key for educational policies promoting computational literacy among high school students and teachers.

## Methods

The analysis presented here is based on data collected by Stella et al. (2019), which included conceptual associations and positive, negative and neutral labels provided by 159 Italian high school students in their final school year and by 59 international STEM researchers. Students from three different Italian high schools located in Southern Italy were involved independently in the data gathering on their school grades. STEM researchers had extensive training in modelling, simulating and understanding complex systems. Both samples of students and researchers have the same ratios of male and female participants.

FMNs combine artificial intelligence, cognitive psychology and complex systems to explore both explicit/conscious and implicit/subconscious knowledge and emotional perception of individuals or groups of individuals toward a given topic (see Stella et al. 2019; Stella 2020 and Stella and Zaytseva,

2020). Behavioral forma mentis networks (BFMNs) represent knowledge as a knowledge graph, i.e. a web of concepts interconnected by knowledge-eliciting links. Unlike WordNet or other knowledge graphs, BFMNs use free associations from psycholinguistics in order to link concepts (De Deyne et al. 2016). Patterns of conceptual associations are provided by individuals through a simple form and reflect how people structure their knowledge around specific concepts. BFMNs are different from textual forma mentis networks (TFMNs). The latter reconstruct knowledge from text analysis and hence do not pass through behavioral tasks as BFMNs do. Both BFMNs and TFMNs are forma mentis networks (FMNs), in that they combine conceptual knowledge and affect patterns. BFMNs include subjective norms for sentiment, as concepts are also rated as "positive", "negative" or "neutral" by individuals. Combining knowledge structure and affective patterns, forma mentis networks identify how concepts are perceived in a given set of individuals or populations. By cross-validation with independent affective norms, words mostly surrounded by negative associates in BFMNs were also found to correspond to anxiety-eliciting concepts, including STEM anxiety (an increased alertness inspired by facing STEM problems) and test anxiety (a feeling of stress/tension felt before or during a test/quiz/exam). For visualisation purposes, in the following plots all Italian words were translated from Italian to English.

## Results

Forma mentis networks give structure to the way concepts are semantically framed and perceived. The following analysis focuses on semantic frames (Fillmore, 2006), which correspond to the semantic associations attributed by individuals to a given concept in the free association game. The sentiment data included in an FMN enable also the identification of positive/negative/neutral perceptions pervading a given semantic frame and thus altering the connotation or aura attributed to a given concept through its associations.

Figure 1 visualises how forma mentis networks give structure to conceptual knowledge and affect in individuals' minds, i.e. a mindset. Figure 1 compares the reconstructed mindset around "model", with high-schoolers associating it mostly with jargon about fashion, role-models and a few scientific terms.

In their mindset, STEM experts strongly associate "model" with computational thinking, research and experimenting.

[HERE FIGURE 1]

Figure 2 reports students' perceptions toward these concepts, including methodological aspects such as "code" and one of the main subjects involved in computing, i.e. "computer science" (for other computational sciences we refer the interested reader to Stella et al. 2019).

Students perceived "computation" as a negative concept, surrounded by a negative emotional aura of negative associates. Negative auras are relative to concentrations of negative valence, which in turn elicit strong anxiety and stress, such as students', "computation"-elicited anxiety. Taking a closer look, this negative aura emerges mainly from algorithmic concepts, the perception of which might be dry and detached from their effective relevance in real-world applications of calculations. This pattern is further indicated by the only positive associations being examples of real-world systems related to calculations, i.e. brain, power, logic and reasoning. Students also provided a rich number of semantic associations to "computer science" and "code", which were distinctly perceived as neutral, and thus different from the mathematical jargon reported in "computation". This sharp emotional difference might indicate that students perceive "code" to be a different entity compared to dry mathematical jargon.

[HERE FIGURE 2]

There are also positive patterns. Students' reconstructed mindsets around "logic reasoning", a key component of computational thinking, features positive semantic associations in relation to understanding the world through reasoning that elicits a sense of openness toward computational thinking. Such positivity was also found in the reconstructed mindset of STEM experts. However, such a positive/open attitude in students directly contrasted with some negative associations featuring dry mathematical jargon and calculation methods.

Computational thinking fundamentally relies on simulations, data and, to a lesser extent, computers (Weintrop et al. 2016). Figure 3 reports students' and researchers' perceptions of these concepts.

For students, "simulation" reminds them only of exams, grades and related anxiety-eliciting concepts (e.g. a simulation of a quiz). This obliviousness to computational concepts indicates a deep lack of awareness about computer simulations, which by comparison are prominently featured in STEM experts' mindset.

[HERE FIGURE 3]

As compared with "code", students perceived "data" as a negative entity, mainly linked to statistics, a concept eliciting statistics anxiety in the sampled student population (see Stella et al. 2019). Notice how in students' reconstructed mindset, "simulation" was not associated with "data", whereas this was a positive and strong link in researchers' networked mindset.

Last but not least, forma mentis networks indicate that students perceived a computer as a technological device for surfing the web and doing calculations, whereas for researchers a computer was an instrument for exploring the science of natural systems through models and simulations. The absence of "model" and "simulation" in the students' mindset suggests a limited perception in students, who see only the computation aspects of computers, i.e. crunching the numbers without understanding their relevance to the real world.

## Discussion

Characterising the cognitive footprint of a computational thinking mindset is urgently needed for devising effective educational policies promoting data literacy skills of relevance for future job prospects (OECD, 2019) and for fighting "functional illiteracy", a concerning inability for literate people to understand quantitative information and data (cf. Vagvolgyi et al. 2016).

Our results show how the newly developed framework of forma mentis networks can tackle this challenge by offering a transparent reconstruction of how individuals perceive knowledge and sentiment about computational literacy. Whereas in STEM experts there was a virtuous circle of positive conceptual associations linking together models, theory, coding and simulations with understanding nature, in high school students this computational mindset was only partly developed.

The investigated population of students possessed advanced training in STEM subjects and was trained in basic numerical simulations. Despite this, students' computational thinking exhibited awareness of the knowledge behind logic reasoning but it identified "computing" as a negative, anxiety-eliciting concept, and completely missed any link between "simulation", "model" and computing itself. Students' semantic frame of "computing" portrayed a perception equivalent to mere number crunching, without associations linking computing to understanding the real world. However, these conceptual associations were prominently detected in the forma mentis network of STEM experts.

Strongly different semantic frames were also found with "model" and "simulation". In researchers, "model" was semantically framed as a key tool for research. In the students' FMN, "model" was semantically framed mostly as related to role-models or fashion runaways, and it elicited only a few scientific associations. This difference suggests that a lack of computational thinking might be related to a vague, abstract and mostly incomplete awareness of the impact that scientific models have for achieving new knowledge in research. Several studies have reported models as being effective in introducing learning in a new domain and in exploring different problem situations (cf. Milrad 2003). Model-facilitated learning enables students to gradually explore problems from an active perspective. Aided by a model, students can identify the structure of a given problem, making it progressively more complicated and tackling the challenge of determining whether predicted outcomes occur. In this way, models can provide a learning experience more similar to actual research than that usually found in mainstream classroom teaching and can be beneficial for fostering computational thinking. Recent approaches such as the one by Sabitzer and colleagues (2020) identified modelling as a key approach for learning computational skills in primary and secondary school, integrating computational thinking with everyday situations and challenges.

Models enable the exploration of how different choices or initial conditions influence a certain problem, mainly through computer simulations. The concept of "simulation" features the most dramatic differences between students' and researchers' FMNs. No association reported in the students' FMN framed simulation as a computer experiment. Instead, "simulation" for students was associated mainly with tests and quizzes and heavily subject to test anxiety (cf. Stella 2020), an abnormally high level of distress perceived before or during exams that critically impairs academic performance. We argue that the utter lack of any computational association with "simulation" is an important indicator of how students are learning logic reasoning and mathematical skills without learning the importance of simulating processes or problems. A crippled or missing awareness of simulations is highly problematic for STEM learning. The *principle of problem centeredness* in science learning clearly indicates that learning can be more effective when learners are engaged in solving real-world problems (Merrill 2002). Since simulations implement an exploration of real-world problems that cannot be often achieved by theoretical formulas or equations, preventing students from learning about simulations (and modelling) can greatly decrease the effectiveness of science learning and thus hamper the development of computational thinking. Notice that this problem afflicted students' mindset but not researchers'. The additional university training undertaken by the latter evidently promoted the knowledge necessary for bringing simulations, models and logic reasoning together as ways for exploring and understanding real-world problems, leading to the researchers' virtuous mindset as represented in their forma mentis network.

All in all, our cognitive map indicates an incomplete computational thinking mindset in students, lacking conceptual links between computational tools and knowledge discovery that here were identified in researchers, instead. The absence of these links suggests a lack of motivation or "sense of purpose" in the way students perceive computation. Let us briefly focus on this (lack of) motivation. In cognitive psychology, the concept of "sense of purpose" mostly relates to personal development and indicates an intention to explore and pursue meaningful achievements that bring satisfaction and fulfillment (Maslow 1970; Neto 2015). In other words, a sense of purpose represents a motivation that drives people to engage with and act upon their surroundings in order to achieve something meaningful, productive and satisfying. Notice that even achieving a clearer understanding of how things work represents a satisfying achievement. Maslow's *theory of motivation* clearly indicates that stimulating a sense of purpose in students is one of the most powerful ways for enhancing learning and skill retention (Neto 2015). Endowing students with a sense of purpose in the context of computational learning means channelling students' computational skills toward curiosity for achieving knowledge about their surrounding world. In this way knowledge represents grounding for computational curiosity, and the combination of the two can give rise to a fully formed computational thinking mindset. Although the methodology of forma mentis networks cannot establish a causal link, the observed lack of motivation could be a cause for the detected negative perception afflicting "computation" and other computing concepts in the students' mindset. In fact, several studies have shown that a lack of a "sense of purpose" can give rise to negative perceptions promoting boredom, negativity and even more extreme consequences such as anxiety (cf. Falham et al. 2009). Analogous psychological mechanisms might be at work here. Without a sense of purpose that puts the mathematical jargon into the perspective of exploring the real world, the semantic frame around "computation" would stagnate in abstract concepts, acquiring a negative, boring and dry connotation (Ashcraft 2002; Ramirez et al. 2012). This potential link calls for future research aimed at identifying a causal relationship between computational thinking development and negative attitudes toward STEM.

It is important to boost positive attitudes toward computation. In fact, several studies indicate that positive attitudes towards domain knowledge are related to better academic performance (Beurze et al. 2013; Chiesi & Primi 2010; Emmioglu and Capa-Aydin 2012). This study supports this relationship by comparing respondents who have different levels of expertise, i.e. high schoolers and researchers. In both cases, the investigation of the perception of key concepts by students and experts is important for establishing the role of attitudes in learning. Attitudes are highly related to individual traits and are rarely subject to drastic changes during learning (Beurze et al. 2013), making it extremely difficult to reduce or relieve negative perceptions of topics at later learning stages. This puts an emphasis on the importance of promoting computational thinking by eliciting positive, meaningful and purposeful framings of computational skills during the early stages of science learning, e.g. during primary school.

# Limitations

It has to be stated that the above analysis suffers from some limitations. The most prominent one is related to the necessity of involving individuals in a cognitive lab experiment, which translates into smaller population sizes, especially for researchers. The recent development of textual forma mentis networks, meaning extracting mindsets from texts, could potentially address this limitation while enabling the analysis of large written corpora extracted from educational forums or textbooks. With the possibility of improving learners' experiences through digitally collected data from social forums (cf. Poquet et al. 2020), the detection of computational thinking patterns starting from text would represent an interesting direction for future research (see also next section). Another limitation of this study is that it measured only semantic frames and sentiment patterns but it did not measure the concrete proficiency of students in computational tasks such as coding or simulating models. Exploring the correlation between negative/positive attitudes toward computational thinking and both coding and simulating models stands as an important direction for future work.

# Conclusions and Next Steps

To create an invitation to more diverse and equitable participation in computer science and participation in the data-driven workforce, there is an urgent need to contextualize and make computational thinking meaningful and purposeful to students throughout their scholastic career in order to have an impact on attitudes toward computer science. Helping learners bring meaning to the use of computers, coding, modeling and data by providing them with opportunities to use these techniques to address problems they care about and/or are important to their communities and society is going to be essential to overcoming the stigma of computer science and related subjects in and out of school. Forma mentis networks will allow controlled trials to get a sense of the impact of meaningful computation programs with students and teachers and can guide educators, administrators, policymakers and lifelong learners in creating a computational thinking future without fear.

As the COVID-19 pandemic demonstrates, the need to understand how data are used and interpreted by researchers and public health officials and are used to inform decisions made at all levels, from personal to policy, is essential to the security and wellbeing of all. And to achieve this understanding, there must be a requisite level of computational thinking. Data science - the science of how scientific, social and economic data are acquired, organized, visualized and analyzed - is revolutionizing all sectors of society while simultaneously affecting equity in the form of data haves and have-nots. The lack of computational skills, computational thinking, and skills in the use of the tools of data science (technologies to gather data from sources such as surveys, sensors, loggers, social media, GPS, open data sources and economic transactions, and the software and computer systems to analyze and visualize them), and the lack of understanding issues of equity, security, and privacy (how and why personal and social data are gathered, analyzed and distributed), impacts equity in employment and the innovation economy.

## Acknowledgements

The authors acknowledge Aleksandra Aloric for providing valuable feedback on this manuscript.

**Figures:**

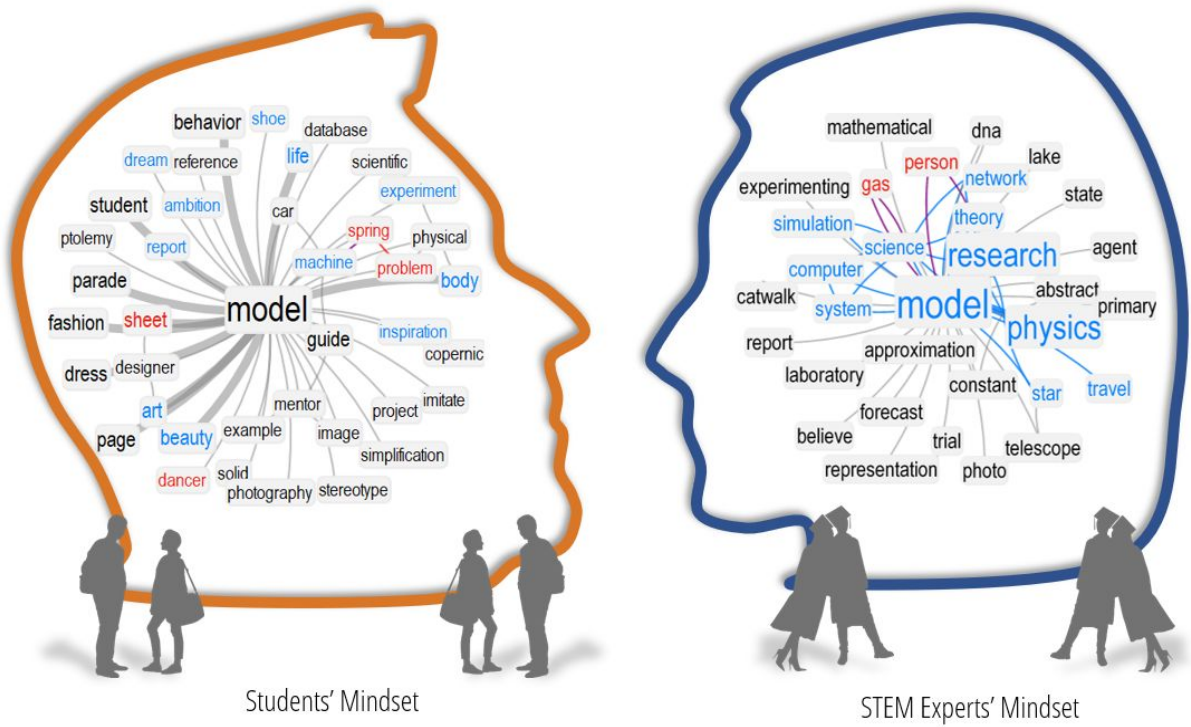

**Figure 1**: Forma mentis networks capture and reconstruct mindsets of individuals from different groups. Visual comparison of high school students' and STEM experts' mindsets around "model". Conceptual links indicate memory recall, e.g. reading "model" made students think of "fashion". Associations provided by two or more individuals are thicker. Words perceived by students as positive (negative) are highlighted in cyan (red). Links between positive (positive/negative) words are highlighted in cyan (purple). Larger font-size indicates higher closeness centrality of a concept.

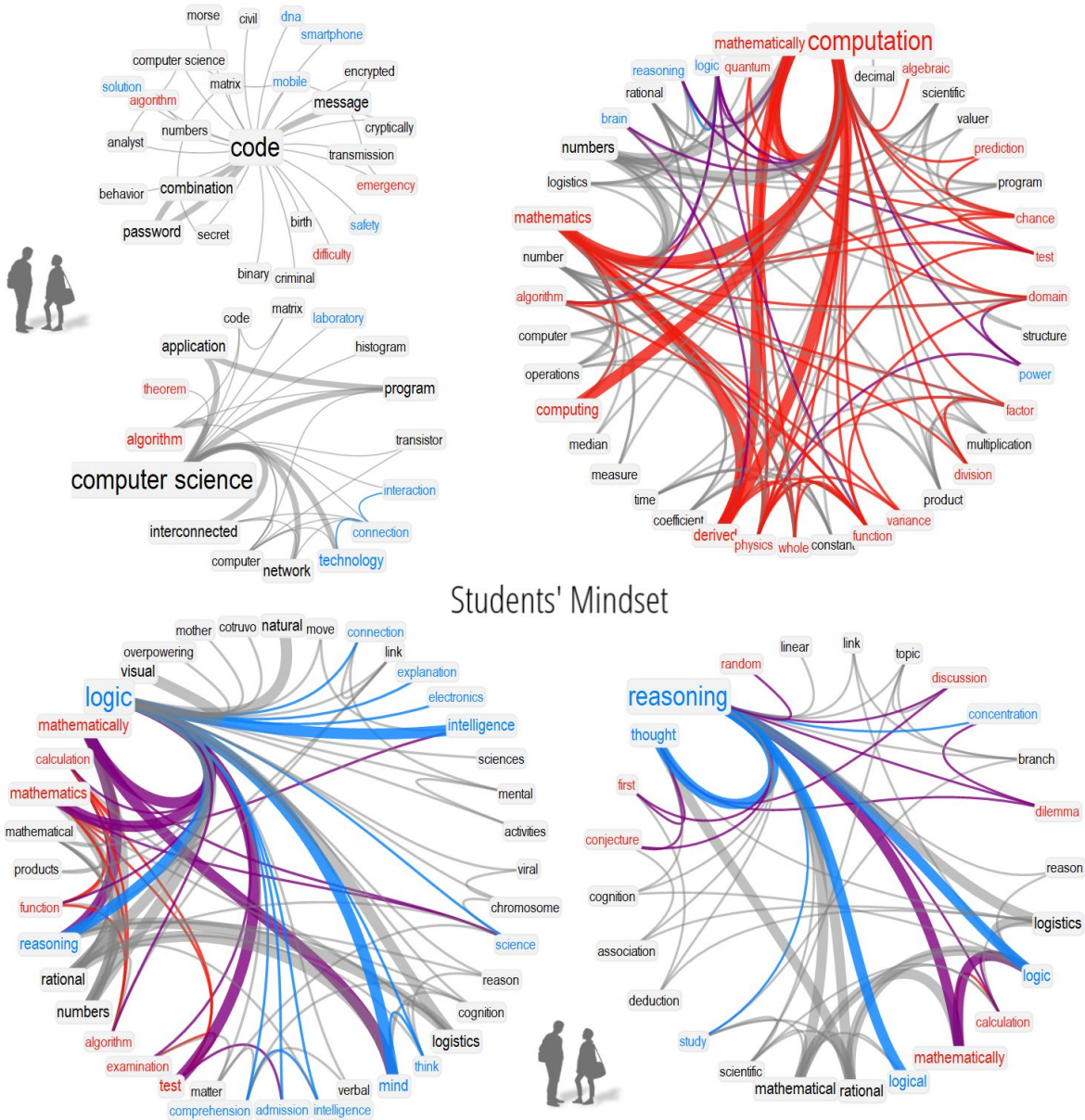

**Figure 2**: Students' mindset around prominent data literacy concepts such as "code", "computer science", "computation", "logic" and "reasoning". Notice how "computation" is surrounded by mostly negatively-perceived concepts, i.e. "computation" elicits a negative emotional aura. Negative concepts are prominently featured also in the students' mindset around "logic" and they consistently deal with mathematics or mathematical jargon.

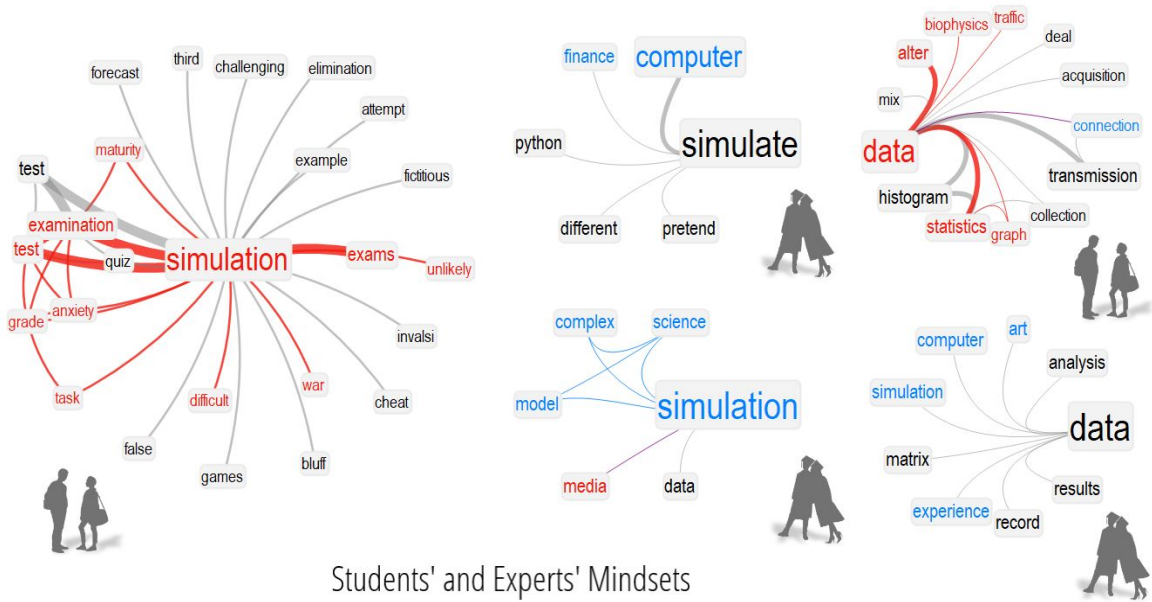
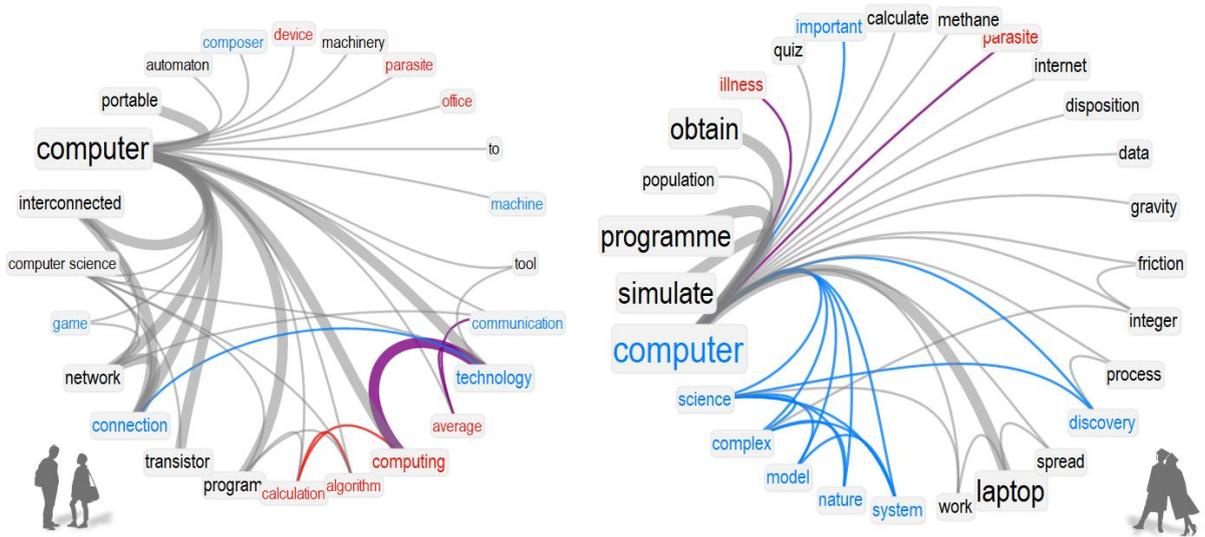

**Figure 3**: Reconstructed mindsets for students and STEM experts around "simulation"/"simulate", "data" and "computer".